\documentclass[letterpaper, 10 pt, conference]{ieeeconf}  

\IEEEoverridecommandlockouts                              

\title{\LARGE \bf Reachability Analysis of Nonlinear Discrete-Time Systems Using Polyhedral Relaxations and Constrained Zonotopes*
}

\author{Brenner S. Rego$^{1,2}$, Guilherme V. Raffo$^{3}$, Marco H. Terra$^{2}$, and Joseph K. Scott$^{1}$
\thanks{*This work was partially supported by the Brazilian agencies CNPq, under grants 465755/2014-3 (INCT project), 317058/2023-1 and 422143/2023-5; FAPESP, under grants 2014/50851-0, 2022/05052-8, and 2023/06896-8; and CAPES through the Academic Excellence Program (PROEX).}
\thanks{$^{1}$Brenner S. Rego and Joseph K. Scott are with the Department of Chemical and Biomolecular Engineering, Georgia Institute of Technology, 311 Ferst Dr., Atlanta, 30318, GA, USA. {\tt\small brego6@gatech.edu, joseph.scott@chbe.gatech.edu}
        }%
\thanks{$^{2}$Brenner S. Rego and Marco H. Terra are with the Department of Electrical and Computer Engineering, University of São Paulo, São Carlos, SP 13566-590, Brazil. {\tt\small brennersr7@usp.br, terra@sc.usp.br}
        }%
\thanks{$^{3}$Guilherme V. Raffo is with the Department of Electronics Engineering and the Graduate Program in Electrical Engineering, Federal University of Minas Gerais, Belo Horizonte, MG 31270-901, Brazil. {\tt\small raffo@ufmg.br}
        }%
}

\usepackage[american]{babel}
\usepackage[utf8]{inputenc}
\usepackage[T1]{fontenc}
\usepackage{graphicx}
\usepackage{amsmath}
\usepackage{amssymb}
\usepackage{amsfonts}
\usepackage{bm}
\usepackage{hhline}
\usepackage{csquotes}
\usepackage{xcolor}
\usepackage{import}
\usepackage{algorithm}
\usepackage{algpseudocode}
\usepackage{dsfont}
\usepackage{multirow}

\usepackage[caption = false]{subfig}
\usepackage{soul}

\newtheorem{proposition}{\bf{Proposition}}
\newtheorem{definition}{\bf{Definition}}
\newtheorem{remark}{\bf{Remark}}

\allowdisplaybreaks 

\newcommand{\mbf}[1]{\ensuremath{{\mathbf{#1}}}}

\newcommand{\half}{\ensuremath{\frac{1}{2}}}

\newcommand{\eye}[1]{\ensuremath{\mbf{I}_{#1}}}

\newcommand{\zeros}[2]{\ensuremath{\bm{0}_{#1\times#2}}}
\newcommand{\ones}[2]{\ensuremath{\bm{1}_{#1\times#2}}}

\newcommand{\real}[1]{\ensuremath{\text{Re}(#1)}}

\newcommand{\realset}{\ensuremath{\mathbb{R}}}
\newcommand{\realsetmat}[2]{\ensuremath{\mathbb{R}^{#1\times#2}}}
\newcommand{\intvalset}{\ensuremath{\mathbb{I}\mathbb{R}}}

\newcommand{\naturalset}{\ensuremath{\mathbb{N}}}

\newcommand{\lbound}{\ensuremath{\text{L}}}
\newcommand{\ubound}{\ensuremath{\text{U}}}
\newcommand{\midpoint}{\ensuremath{\text{M}}}

\newcommand{\radius}{\ensuremath{\text{R}}}

\newcommand{\lb}[1]{{#1}^\lbound}
\newcommand{\ub}[1]{{#1}^\ubound}
\newcommand{\midp}[1]{{#1}^\midpoint}
\newcommand{\rad}[1]{{#1}^\radius}

\newcommand{\poly}{\ensuremath{_\text{P}}}
\newcommand{\zon}{\ensuremath{_\text{Z}}}
\newcommand{\czon}{\ensuremath{_\text{CZ}}}

\newcommand{\intval}[1]{\ensuremath{[\lb{#1}, \ub{#1}]}}

\newcommand{\ninf}[1]{\ensuremath{\|{#1}\|_\infty}}

\newcommand{\noarg}{\ensuremath{\_\,}}

\newcommand{\bibfolder}{Bibliography}

\begin{document}

\maketitle
\thispagestyle{empty}
\pagestyle{empty}

\begin{abstract}
This paper presents a novel algorithm for reachability analysis of nonlinear discrete-time systems. The proposed method combines constrained zonotopes (CZs) with polyhedral relaxations of factorable representations of nonlinear functions to propagate CZs through nonlinear functions, which is normally done using conservative linearization techniques. The new propagation method provides better approximations than those resulting from linearization procedures, leading to significant improvements in the computation of reachable sets in comparison to other CZ methods from the literature. Numerical examples highlight the advantages of the proposed algorithm.
\end{abstract}


\section{Introduction}
Set-based computations have become important in many fields of research in recent decades. Applications include fault detection and diagnosis \cite{Raimondo2016,Zhang2023}, state and parameter estimation \cite{Rego2022Joint}, robot localization \cite{Rohou2020,Bhamidipatinavi2022}, nonlinear model predictive control with obstacle avoidance \cite{Nascimento2023}, tube-based model predictive control \cite{Richard2024MPCCZ}, and reachability analysis \cite{Althoff2021Reach}. Set-based methods are able to generate guaranteed enclosures in applications affected by unknown-but-bounded uncertainties, unlike stochastic strategies which require knowledge of the stochastic properties of the uncertainties \cite{Simon2010}.  

Reachability analysis of dynamical systems is a very important topic in set-based computating. The reachability problem consists of obtaining guaranteed enclosures of the system states at future times given enclosures of the initial states and any sort of uncertainties. Such computations require algorithms for propagating sets through the system dynamics, which are generally nonlinear. The enclosures may be used to predict the behavior of the system for different initial states and parameters, or even to evaluate if the mathematical model is accurate by comparing with measurements from a real plant.

For linear systems, accurate enclosures can be efficiently computed using zonotope methods due to the computational advantages of these sets for some important operations, such as the Minkowski sum \cite{Serry2022}. Nevertheless, the computation of good enclosures for nonlinear functions, and, consequently, accurate reachability analysis of nonlinear systems, is still an open problem \cite{Alamo2005a,Rego2021}. Several set representations have been used in this context, such as convex polytopes \cite{Shamma1997}, intervals \cite{Jaulin2016}, zonotopes \cite{Alamo2005a,Combastel2005}, and generalizations of zonotopes, such as constrained zonotopes (CZs) \cite{Rego2021}, constrained polynomial zonotopes \cite{Kochdumper2023CPZ}, and hybrid zonotopes \cite{Siefert2024HZFuncDecomp}. In this work, we focus on methods based on zonotopes and constrained zonotopes, since well-established and efficient complexity reduction algorithms are available  \cite{Scott2016,Scott2018,Raghuraman2022}.

Existing nonlinear reachability methods based on zonotopes and CZs often rely on a conservative linear approximation of the right-hand side function, with bounds on the linearization error obtained from either the Mean Value Theorem or Taylor's Theorem \cite{Alamo2005a,Rego2021,Combastel2005}. These approaches can provide accurate enclosures in cases where the uncertainties are small enough that the true reachable set never becomes too large. In such situations, the linearization is done over a modest domain in each step and can be quite accurate. However, if the true reachable sets become large, then the linearization step can generate severe conservatism, even if the current enclosure is very accurate. This proves to be a major drawback in many applications, rendering the computed enclosures unusable. Improved bounds on the linearization error can be obtained in some cases using DC programming \cite{Alamo2008,DePaula2024DC}. However, these algorithms have exponential complexity and the results may not be accurate for certain functional forms \cite{Tottoli2023}.

In this paper, we present a new algorithm for reachability analysis of nonlinear discrete-time systems that uses lifted polyhedral relaxations to propagate constrained zonotopes through nonlinear functions. Constrained zonotopes \cite{Scott2016} are an extension of zonotopes capable of describing convex polytopes with arbitrary complexity while retaining most of the computational advantages of zonotopes \cite{Kuhn1998}. In the new propagation method, the nonlinear right-hand side function is decomposed into a sequence of elementary operations, known as a \emph{factorable representation}. This representation is then used to generate a polyhedral enclosure in an augmented space. This approach is commonly used in the global optimization community for constructing linear programming relaxations of nonlinear optimization problems \cite{Tawarmalani2002Convexification,Tawarmalani2005Polyhedral}. Properties of CZs are then employed to efficiently project the enclosure into the function's image space. The resulting enclosure is a CZ, thus allowing the recursive computation of CZ enclosures for reachability analysis. This propagation method causes only a linear increase in the complexity of the enclosures, which is easily handled using well-known complexity reduction algorithms for CZs \cite{Scott2016,Raghuraman2022}. Computational results indicate that this approach offers significant advantages over existing CZ methods based on conservative linearization techniques \cite{Rego2021}.

\subsection*{Notation}

Lowercase italic letters denote scalars, lowercase bold letters denote vectors, uppercase bold letters denote matrices, and uppercase italic letters denote general sets. The sets of natural numbers and real numbers are denoted by $\naturalset$ and $\realset$, respectively. Moreover, $\zeros{n}{m}$ and $\ones{n}{m}$ denote $n \times m$ matrices of zeros and ones, respectively.

\section{Problem Statement and Preliminaries} \label{sec:probformulation}

Consider a class of discrete-time systems with nonlinear dynamics described by
\begin{equation}
	\mbf{x}_k = \mbf{f}(\mbf{x}_{k-1}, \mbf{w}_{k-1}), \label{eq:systemf}   
\end{equation} 
where $\mbf{x}_k \in \realset^{n_x}$ is the system state and $\mbf{w}_k \in \realset^{n_w}$ is the process uncertainty. The initial condition and uncertainty are assumed to be bounded, meaning that $\mbf{x}_0 \in X_0$ and $\mbf{w}_{k} \in W$, $\forall k \geq 0$, where $X_0$ and $W$ are known convex sets.%

The objective of this work is to obtain convex enclosures $\bar{X}_k$ of the system states $\mbf{x}_k$ for any $k\geq 0$ as accurately as possible. With $\bar{X}_0 \triangleq X_0$, this is accomplished using the recursive approach
\begin{equation}
\bar{X}_k \supseteq \{ \mbf{f}(\mbf{x}_{k-1}, \mbf{w}_{k-1}): (\mbf{x}_{k-1}, \mbf{w}_{k-1}) \in \bar{X}_{k-1} \times W\}. \label{eq:reachrecursive}
\end{equation}
With $X_{0}$ and $W$ described as CZs, the key step consists of propagating constrained zonotopes through the nonlinear function $\mbf{f} : \realset^{n_x} \times \realset^{n_w} \to \realset^{n_x}$.

\subsection{Interval analysis}

Let $\intvalset^n$ denote the set of all non-empty compact intervals in $\realset^n$. For \emph{endpoints} $\lb{\mbf{x}}, \ub{\mbf{x}} \in \realset^n$ with $\lb{\mbf{x}} \leq \ub{\mbf{x}}$, an \emph{interval} $X \in \intvalset^n$ is defined as $X \triangleq \{\mbf{x} \in \realset^n: \lb{\mbf{x}} \leq \mbf{x} \leq \ub{\mbf{x}}\} \triangleq \intval{\mbf{x}}$. In addition, $\text{mid}(X) \triangleq \half (\ub{\mbf{x}} + \lb{\mbf{x}}) \triangleq \midp{\mbf{x}}$, $\text{rad}(X) \triangleq \half (\ub{\mbf{x}} - \lb{\mbf{x}}) \triangleq \rad{\mbf{x}}$, and $B_\infty^n \triangleq [-\ones{n}{1}, \ones{n}{1}]$.

Let $X \triangleq \intval{x} \in \intvalset$ and $W \triangleq \intval{w} \in \intvalset$. Then, for any of the four basic arithmetic operations $\odot\in\{+,-,\times,/\}$, we define $X \odot W \triangleq \{ x \odot y : x \in X, w \in W \}$ (division is undefined if $0 \in W$). Moreover, interval extensions of elementary functions $h: \realset \to \realset$, such as the power, exponential, logarithm, and trigonometric functions, are defined as $h: \intvalset \to \intvalset$, $h(X) \triangleq \{h(x):x \in X\}$. Simple formulas for computing all of these operations in terms of the input endpoints can be found in \cite{Moore2009}.

\subsection{Convex polytopes and constrained zonotopes}

\emph{Convex polytopes} are convex sets that can be described as either the intersection of a finite set of halfspaces (H-rep) or the convex hull of a finite set of vertices (V-rep). In this work, we are interested in the former. In the following definition, we extend the typical halfspace representation to explicitly include linear equality constraints, which is a change in notation only since any linear equality can be represented as the intersection of two opposing halfspaces.

\begin{definition} \rm \label{def:hrep}
	A set $P \subset \realset^n$ is a \emph{convex polytope} in halfspace representation if there exists $(\mbf{H}_p, \mbf{k}_p, \mbf{A}_p, \mbf{b}_p) \in \realsetmat{n_h}{n} \times \realset^{n_h} \times \realsetmat{n_{c_p}}{n} \times \realset^{n_{c_p}}$ such that
	\begin{equation} \label{eq:defhrep}
	P = \{ \mbf{x} \in \realset^n : \mbf{H}_p \mbf{x} \leq \mbf{k}_p, \ \mbf{A}_p \mbf{x} = \mbf{b}_p \}.
	\end{equation}	
\end{definition}

\emph{Constrained zonotopes} are an extension of zonotopes \cite{Kuhn1998} that include linear equality constraints. The CZ representation (CZ-rep) defined next is an alternative representation for convex polytopes that retains many of the computational advantages of zonotopes.

\begin{definition} \cite{Scott2016} \rm \label{def:pre_czonotopes}
	A set $Z \subset \realset^n$ is a \emph{constrained zonotope} if there exists $(\mbf{G}_z,\mbf{c}_z,\mbf{A}_z,\mbf{b}_z) \in \realsetmat{n}{n_g} \times \realset^n \times \realsetmat{n_c}{n_g} \times \realset^{n_c}$ such that
	\begin{equation} \label{eq:pre_cgrep}
	Z = \left\{ \mbf{c}_z + \mbf{G}_z \bm{\xi} : \| \bm{\xi} \|_\infty \leq 1, \ \mbf{A}_z \bm{\xi} = \mbf{b}_z \right\}.
	\end{equation}	
\end{definition}

In \eqref{eq:defhrep}, each inequality is a \emph{halfspace}. In \eqref{eq:pre_cgrep}, each column of $\mbf{G}_z$ is a \emph{generator}, $\mbf{c}_z$ is the \emph{center}, and $\mbf{A}_z \bm{\xi} = \mbf{b}_z$ are the \emph{constraints}. We use the compact notation $P = (\mbf{H}_p, \mbf{k}_p, \mbf{A}_p, \mbf{b}_p)\poly$ for convex polytopes in H-rep, $Z = (\mbf{G}_z, \mbf{c}_z,\mbf{A}_z,\mbf{b}_z)\czon$ for CZs, and $Z = (\mbf{G}_z, \mbf{c}_z)\zon$ for zonotopes. The latter two are referred to as the constrained generator representation (CG-rep) and generator representation (G-rep) of these sets, respectively. Moreover, $(\mbf{H}_p, \mbf{k}_p, \noarg, \noarg)\poly$ and $(\noarg, \noarg, \mbf{A}_p, \mbf{b}_p)\poly$ denote polytopes with only inequality constraints and equality constraints, respectively. It is noteworthy that an interval $X \in \intvalset^n$ can be described in G-rep as $(\text{diag}(\text{rad}(X)),\text{mid}(X))\zon$.

Consider sets $Z, W \subset \realset^{n}$, $Y \subset \realset^{m}$, and a matrix $\mbf{R} \in \realset^{m \times n}$. Define the Cartesian product, linear image, Minkowski sum, and generalized intersection, as $Z \times W \triangleq \{(\mbf{z},\mbf{w}): \mbf{z} \in Z, \mbf{w} \in W\}$, $\mbf{R}Z  \triangleq \{ \mbf{R} \mbf{z} : \mbf{z} \in Z\}$, $Z \oplus W  \triangleq \{ \mbf{z} + \mbf{w} : \mbf{z} \in Z,\, \mbf{w} \in W\}$, and $Z \cap_{\mbf{R}} Y  \triangleq \{ \mbf{z} \in Z : \mbf{R} \mbf{z} \in Y\}$, respectively. If $Z \triangleq (\mbf{G}_z, \mbf{c}_z, \mbf{A}_z, \mbf{b}_z)\czon \subset \realset^n$, $W \triangleq (\mbf{G}_w, \mbf{c}_w, \mbf{A}_w, $ $\mbf{b}_w)\czon \subset \realset^n$, and $Y \triangleq (\mbf{G}_y, \mbf{c}_y, \mbf{A}_y, \mbf{b}_y)\czon \subset \realset^m$ are constrained zonotopes, then 
\begin{align}
Z {\times} W & = \! \left( \begin{bmatrix} \mbf{G}_z  & \bm{0} \\ \bm{0} & \mbf{G}_w \end{bmatrix}\!, \begin{bmatrix} \mbf{c}_z \\ \mbf{c}_w \end{bmatrix}\!, \begin{bmatrix} \mbf{A}_z & \bm{0} \\ \bm{0} & \mbf{A}_w \end{bmatrix}\!, \begin{bmatrix} \mbf{b}_z \\ \mbf{b}_w \end{bmatrix} \right)\czon\!\!\!\!, \label{eq:pre_czcartprod}\\
\mbf{R}Z & = \left( \mbf{R} \mbf{G}_z, \mbf{R} \mbf{c}_z, \mbf{A}_z, \mbf{b}_z \right)\czon, \label{eq:pre_czlimage}\\
Z {\oplus} W & = \left( [ \mbf{G}_z \,\; \mbf{G}_w ], \mbf{c}_z + \mbf{c}_w, \begin{bmatrix} \mbf{A}_z & \bm{0} \\ \bm{0} & \mbf{A}_w \end{bmatrix}\!, \begin{bmatrix} \mbf{b}_z \\ \mbf{b}_w \end{bmatrix} \right)\czon\!\!\!\!, \label{eq:pre_czmsum}\\
Z {\cap_{\mbf{R}}} Y & = \left( [\mbf{G}_z \,\; \bm{0}], \mbf{c}_z, \begin{bmatrix} \mbf{A}_z & \bm{0} \\ \bm{0} & \mbf{A}_y \\ \mbf{R} \mbf{G}_z & -\mbf{G}_y \end{bmatrix}, \begin{bmatrix} \mbf{b}_z \\ \mbf{b}_y \\ \mbf{c}_y - \mbf{R} \mbf{c}_z \end{bmatrix} \right)\czon\!\!\!\!. \label{eq:pre_czintersection}
\end{align}

Define $B_\infty(\mbf{A}_z,\mbf{b}_z) \triangleq \{\bm{\xi} \in \realset^{n_g} : \ninf{\bm{\xi}} \leq 1,\,  \mbf{A}_z \bm{\xi} = \mbf{b}_z \}$. Then, $(\mbf{G}_z, \mbf{c}_z,\mbf{A}_z,\mbf{b}_z)\czon = \mbf{c}_z \oplus \mbf{G}_z B_\infty(\mbf{A}_z,\mbf{b}_z)$ holds, and $(\mbf{G}_z, \mbf{c}_z)\zon = \mbf{c}_z \oplus \mbf{G}_z B_\infty^n$. Additionally, if $P \triangleq (\mbf{H}_p, \mbf{k}_p, \mbf{A}_p, \mbf{b}_p)\poly \subset \realset^n$ and $Q \triangleq (\mbf{H}_q, \mbf{k}_q, \mbf{A}_q, \mbf{b}_q)\poly \subset \realset^n$, then
\begin{equation} \label{eq:hrepintersection}
P \cap Q = \left( \begin{bmatrix} \mbf{H}_p \\ \mbf{H}_q \end{bmatrix}, \begin{bmatrix} \mbf{k}_p \\ \mbf{k}_q \end{bmatrix}, \begin{bmatrix} \mbf{A}_p \\ \mbf{A}_q \end{bmatrix}, \begin{bmatrix} \mbf{b}_p \\ \mbf{b}_q \end{bmatrix} \right)\poly.
\end{equation}

Efficient methods to enclose a CZ by another one with fewer generators and constraints are available in \cite{Scott2016}. Moreover, the interval hull of a CZ, $Z \subset \realset^n$, denoted as $\square Z \in \intvalset^n$, can be computed by solving $2n$ linear programs (LPs) \cite{Scott2016,Rego2018}.

\section{Lifted Halfspace Polyhedral Enclosure of a Nonlinear Function over an Interval} \label{sec:hrepenclosureinterval}

In this paper, we consider nonlinear functions that are \emph{factorable} as defined below. This definition refers to a library $\mathcal{L}$ of \emph{intrinsic univariate functions}, which typically contains the functions in a standard math library in any programming language, such as $x^a$, $e^x$, $\ln(x)$, $\sin{x}$, etc.

\begin{definition} \label{def:factorable} \rm A function $\mbf{h}: \realset^n \to \realset^m$ is said to be \emph{factorable} if it can be expressed in terms of a finite number of factors $\mbf{z}=(z_1,\ldots,z_{n_z})$ such that, given $\mbf{x} \in \realset^n$,
\begin{enumerate}
    \item $z_j = x_j$ for all $j \in \{1,\ldots,n\}$,
    \item for each $j>n$, $z_j=g_j(\mbf{z})$, where either
        \begin{enumerate}
            \item $g_j(\mbf{z}) \triangleq z_a\odot z_b$ with $a,b < j$ and $\odot\in\{+,-,\times,/\}$, or
            \item $g_j(\mbf{z}) \triangleq \beta_j(z_a)$ with $a < j$ and $\beta_j$ an intrinsic univariate function in $\mathcal{L}$,
        \end{enumerate}
    \item $\mbf{h}(\mbf{x}) = \mbf{E}_h \mbf{z}$, where $\mbf{E}_h \in \realsetmat{m}{n_z}$ is a matrix of zeros except for a single $1$ in each of its rows (i.e., each output of $\mbf{h}$ is an element of $\mbf{z}$).
\end{enumerate}
\end{definition}

For illustration purposes, let $h(\mbf{x}) \triangleq \frac{e^{x_1}}{x_2^2x_3}$. One possible factorable representation for this function is: $z_1 \triangleq x_1$, $z_2 \triangleq x_2$, $z_3 \triangleq x_3$, $z_4 \triangleq g_{4}(\mbf{z}) \triangleq e^{z_1}$, $z_5 \triangleq g_{5}(\mbf{z}) \triangleq z_2^2$, $z_6 \triangleq g_{6}(\mbf{z}) \triangleq z_5 z_3$, and $z_7 \triangleq g_{7}(\mbf{z}) \triangleq \frac{z_4}{z_6}$. %
Then, $h(\mbf{x}) = [\zeros{1}{6} \,\; 1]\mbf{z}$, so $h$ is factorable. Note that the factorization of $h(\mbf{x})$ is not unique. Assuming factorability is not very restrictive, as any function that can be explicitly written in computer code using a standard math library is factorable.

Given an interval $X \in \intvalset^n$ and a factorable function $\mbf{h}: \realset^n \to \realset^m$, the objective of this section is to compute a polyhedral enclosure $P \in \realset^{n_z}$ in the lifted space of the factors $\mbf{z}$ satisfying $\{\mbf{h}(\mbf{x}) : \mbf{x} \in X\} \subseteq \{\mbf{E}_h \mbf{z}: \mbf{z} \in P\}$. To begin, we first compute an interval $Z \in \intvalset^{n_z}$ enclosing all $\mbf{z}$ by recursively taking the natural interval extension of each $z_j = g_{j}(\mbf{z})$ \cite{Moore2009}. Next, for each factor $z_j$ with $j>n$, we derive a convex polytope $Q_j \subset \realset^{n_z}$ in H-rep such that $Q_j \supseteq \{\mbf{z} \in Z : z_j = g_{j}(\mbf{z})\}$. Finally, we define
\begin{equation} \label{eq:liftedpolyhedron}
    P \triangleq \bigcap_{j=n+1}^{n_z} Q_j.
\end{equation}
This set contains all possible $\mbf{z}$ corresponding to $\mbf{x}\in X$. Thus, for any $\mbf{x}\in X$, there must exist $\mbf{z}\in P$ such that $\mbf{h}(\mbf{x}) = \mbf{E}_h \mbf{z}$. It follows that $\{\mbf{h}(\mbf{x}) : \mbf{x} \in X\} \subseteq \{\mbf{E}_h \mbf{z}: \mbf{z} \in P\}$ as desired.

The following subsections show how to obtain enclosures $Q_j$ for some elementary operations. Once these are known, the intersection \eqref{eq:liftedpolyhedron} is computed trivially using \eqref{eq:hrepintersection}, resulting in a convex polytope in H-rep. 

\subsection{Arithmetic operations}

\paragraph{Sum}

Let $z_j = z_a + z_b$, with $a,b < j$, $j > n$. An exact enclosure is given by $Q_j = (\noarg,\noarg, \mbf{r}_+, 0)\poly$, where $\mbf{r}_+$ is a row vector of zeros, except for the $a$th, $b$th, and $j$th columns, which are $1$, $1$, and $-1$, respectively.

\paragraph{Subtraction}

Let $z_j = z_a - z_b$, with $a,b < j$, $j > n$. An exact enclosure is given by $Q_j = (\noarg,\noarg, \mbf{r}_-, 0)\poly$, where $\mbf{r}_-$ is a row vector of zeros, except for the $a$th, $b$th, and $j$th columns, which are $1$, $-1$, and $-1$, respectively.

\paragraph{Multiplication}

Let $z_j = z_a z_b$, with $a,b < j$, $j > n$, $z_a \in [\lb{z}_a, \ub{z}_a]$, and $z_b \in [\lb{z}_b, \ub{z}_b]$. The halfspace enclosure $Q_j \triangleq (\mbf{R}_*, \mbf{s}_*, \noarg, \noarg)$ is derived by rearranging the four inequalities obtained from the McCormick envelope of the bilinear function $z_j = z_a z_b$, which are
\begin{align*}
    z_j & \geq \lb{z}_a z_b + z_a \lb{z}_b - \lb{z}_a \lb{z}_b, \\
    z_j & \geq \ub{z}_a z_b + z_a \ub{z}_b - \ub{z}_a \ub{z}_b, \\
    z_j & \leq \lb{z}_a z_b + z_a \ub{z}_b - \lb{z}_a \ub{z}_b, \\
    z_j & \leq \ub{z}_a z_b + z_a \lb{z}_b - \ub{z}_a \lb{z}_b.
\end{align*}

\paragraph{Division}

Let $z_j = \frac{z_a}{z_b}$, with $a,b < j$, $j > n$, $z_a \in [\lb{z}_a, \ub{z}_a]$, and $z_b \in [\lb{z}_b, \ub{z}_b]$. The halfspace enclosure $Q_j \triangleq (\mbf{R}_/, \mbf{s}_/, \noarg, \noarg)$, is derived by rewriting $z_j = \frac{z_a}{z_b}$ as $z_a = z_b z_j$ and applying the multiplication enclosure accordingly.

\begin{remark} \rm \label{rem:constantoperations}
    For implementation purposes, simpler H-rep descriptions $Q_j$ can be obtained for the case of arithmetic operations with constants, such as $v_j = qv_a$, with $q \in \realset$. %
\end{remark}

\subsection{Univariate functions}

Let $z_j = \beta_j(z_a)$, where $a < j$, $j > n$, $z_a \in Z_a \triangleq [\lb{z}_a, \ub{z}_a]$, and $\beta_j$ is an intrinsic univariate function in the library $\mathcal{L}$. For all such functions, we assume that convex and concave relaxations on $Z_a$ can be readily constructed. Specifically, given any $Z_a\in\mathbb{IR}$, we have convex and concave functions $\beta_j^\text{CV} : Z_a \to \realset$ and $\beta_j^\text{CC} : Z_a \to \realset$, respectively, such that
\begin{equation}
\label{Eq: Beta relaxations bound}
    \beta_j^\text{CV}(z_a) \leq \beta_j(z_a) \leq \beta_j^\text{CC}(z_a), \quad \forall z_a \in Z_a.
\end{equation}
Such relaxations are tabulated for a wide variety of common univariate functions in many global optimization references; see e.g.~Chapter 2 in \cite{ScottPhdThesis}.

Using these functions, we seek to compute a polyhedral enclosure of the form $Q_j = Q^\text{CV}_j \cap Q^\text{CC}_j$, where
\begin{alignat}{1}
\label{Eq: QCV inclusion}
Q^\text{CV}_j &\supseteq \{\mbf{z} \in \realset^{n_z} : z_j \geq \beta_j^\text{CV}(z_a), ~ z_a \in Z_a\}, \\
\label{Eq: QCC inclusion}
Q^\text{CC}_j &\supseteq \{\mbf{z} \in \realset^{n_z} : z_j \leq \beta_j^\text{CC}(z_a), ~ z_a \in Z_a\}.
\end{alignat}
Then, it holds that $Q_j \supseteq \{\mbf{z} \in \realset^{n_z} : z_j = \beta_{j}(z_a), ~ z_a \in Z_a\}$, as desired. Since $\beta_j^\text{CV}$ is convex and $\beta_j^\text{CC}$ is concave, the inequalities in \eqref{Eq: QCV inclusion}--\eqref{Eq: QCC inclusion} remain true if $\beta_j^\text{CV}$ and $\beta_j^\text{CC}$ are replaced by their linearizations at any point in $Z_a$. Therefore, our general strategy is to define $Q^\text{CV}_j$ and $Q^\text{CC}_j$ in terms of linearizations of $\beta_j^\text{CV}$ and $\beta_j^\text{CC}$ at a set of reference points. In many cases, these functions are linear, so no linearization is needed. Otherwise, we use linearizations at $\lb{z}_a$, $\midp{z}_a$, and $\ub{z}_a$. A few specific examples are given below.

\paragraph{Exponential}

$z_j = e^{z_a}$. In this case, $\beta_{j}(z_a)$ is convex, so $\beta^\text{CV}_{j}=\beta_{j}$. Thus, $Q^\text{CV}_j$ is obtained by linearizing $\beta_{j}$ at $\lb{z}_a$, $\midp{z}_a$, and $\ub{z}_a$, leading to the inequalities $z_j  \geq e^{\lb{z}_a} (z_a - \lb{z}_a) + e^{\lb{z}_a}$, $z_j  \geq e^{\midp{z}_a} (z_a - \midp{z}_a) + e^{\midp{z}_a}$, and $z_j  \geq e^{\ub{z}_a} (z_a - \ub{z}_a) + e^{\ub{z}_a}$. The concave relaxation on $Z_a$ is the secant $\beta_j^\text{CC}(z_a) = \left( \frac{e^{\ub{z}_a} - e^{\lb{z}_a}}{\ub{z}_a - \lb{z}_a} \right) (z_a - \lb{z}_a) + e^{\lb{z}_a}$. Thus, $Q^\text{CC}_j$ is defined by the single inequality $z_j  \leq \beta_j^\text{CC}(z_a)$.

\paragraph{Logarithm}

$z_j = \ln(z_a)$.  In this case, $\beta_{j}(z_a)$ is concave, so $\beta^\text{CC}_{j}=\beta_{j}$. Thus, $Q^\text{CC}_j$ is obtained by linearizing $\beta_{j}$ at $\lb{z}_a$, $\midp{z}_a$, and $\ub{z}_a$, leading to $z_j  \leq \frac{1}{\lb{z}_a} (z_a - \lb{z}_a) + \ln(\lb{z}_a)$, $z_j  \leq \frac{1}{\midp{z}_a} (z_a - \midp{z}_a) + \ln(\midp{z}_a)$, and $z_j  \leq \frac{1}{\ub{z}_a} (z_a - \ub{z}_a) + \ln(\ub{z}_a)$. The convex relaxation on $Z_a$ is the secant $\beta_j^\text{CV}(z_a) = \left( \frac{\ln(\ub{z}_a) - \ln(\lb{z}_a)}{\ub{z}_a - \lb{z}_a} \right) (z_a - \lb{z}_a) + \ln(\lb{z}_a)$.  Thus, $Q^\text{CV}_j$ is defined by the single inequality $z_j  \geq \beta_j^\text{CV}(z_a)$.

\paragraph{Even integer power}

$z_j = z_a^{q}$ with $q$ an even integer. In this case, $\beta_{j}(z_a)$ is convex, so $\beta^\text{CV}_{j}=\beta_{j}$. Thus, $Q^\text{CV}_j$ is obtained by linearizing $\beta_{j}$ at $\lb{z}_a$, $\midp{z}_a$, and $\ub{z}_a$, leading to the inequalities $z_j  \geq q(\lb{z}_a)^{(q-1)} (z_a - \lb{z}_a) + (\lb{z}_a)^q$, $z_j  \geq q(\midp{z}_a)^{(q-1)} (z_a - \midp{z}_a) + (\midp{z}_a)^q$, and $z_j  \geq q(\ub{z}_a)^{(q-1)} (z_a - \ub{z}_a) + (\ub{z}_a)^q$. The concave relaxation on $Z_a$ is the secant $\beta_j^\text{CC}(z_a) = \left( \frac{(\ub{z}_a)^q - (\lb{z}_a)^q}{\ub{z}_a - \lb{z}_a} \right) (z_a - \lb{z}_a) + (\lb{z}_a)^q$. Thus, $Q^\text{CC}_j$ is defined by the single inequality $z_j  \leq \beta_j^\text{CC}(z_a)$.

\paragraph{Odd integer power}

$z_j = z_a^{q}$ with $q$ an odd integer. In this case, $\beta_{j}(z_a)$ is concave for $z_a \leq 0$, and convex for $z_a \geq 0$. Therefore, for $\ub{z}_a \leq 0$, the enclosures $Q^\text{CV}_j$ and $Q^\text{CC}_j$ are obtained analogously to the logarithm, while for $\lb{z}_a \geq 0$, the enclosures they are obtained analogously to the even integer power. If $0 \in [\lb{z}_a,\ub{z}_a]$, the procedure is significantly more involved and can be found in Chapter 2 of \cite{ScottPhdThesis}.

\section{Propagation of CZs through Nonlinear Functions using the Halfspace Enclosure} \label{sec:hrepenclosurecz}

Consider a factorable function $\mbf{h}: \realset^n \to \realset^m$. The objective of this section is to compute a constrained zonotope enclosure $H \subset \realset^m$ satisfying $H \supseteq \{\mbf{h}(\mbf{x}) : \mbf{x} \in X\}$, where the domain $X \subset \realset^n$ is now a constrained zonotope.

By using the method developed in Section \ref{sec:hrepenclosureinterval} with interval domain $\square X$, it holds that $\{\mbf{h}(\mbf{x}) : \mbf{x} \in \square X\} \subseteq \{\mbf{E}_h \mbf{z} : \mbf{z} \in P\}$ with $P$ given by \eqref{eq:liftedpolyhedron}. Let $\mbf{z} \triangleq (\mbf{x},\tilde{\mbf{z}})$, with $\tilde{\mbf{z}} \in \tilde{Z}$, where $\tilde{Z} \in \intvalset^{n_z-n}$ is the interval corresponding to the $n_z-n$ bottom rows of $Z \in \intvalset^{n_z}$. The CZ enclosure $H$ is readily obtained by 
\begin{align}
\{\mbf{h}(\mbf{x}) : \mbf{x} \in X\} & \subseteq \{\mbf{E}_h\mbf{z}: \mbf{z} \in X \times \tilde{Z},~ \mbf{z} \in P \}\nonumber\\
& = \mbf{E}_h ((X \times \tilde{Z}) \cap P) \triangleq H. \label{eq:czenclosurefinal}
\end{align}

The Cartesian product in \eqref{eq:czenclosurefinal} is computed by first converting $\tilde{Z}$ into G-rep and then using \eqref{eq:pre_czcartprod}. The multiplication by $\mbf{E}_h$ is computed using \eqref{eq:pre_czlimage}, while the intersection of the constrained zonotope $X \times \tilde{Z}$ and the convex polytope $P$ is obtained by using the following proposition, which is a generalization of the intersection method proposed in \cite{Raghuraman2022}.

\begin{proposition} \rm \label{prop:czhrepintersection}
	Let $Z = (\mbf{G}_z, \mbf{c}_z, \mbf{A}_z, \mbf{b}_z)\czon \subset \realset^n$ be a constrained zonotope and $P \triangleq (\mbf{H}_p, \mbf{k}_p, \mbf{A}_p, \mbf{b}_p)\poly$ be a convex polytope in H-rep with $n_h$ halfspaces. Then, 
	\begin{equation} \label{eq:czhrepintersection}
  \begin{aligned}
		Z \cap P = & \left( [\mbf{G}_z \,\; \mbf{0}], \mbf{c}_z, \left[\begin{smallmatrix} \mbf{A}_z & \mbf{0} \\ \mbf{H}_p \mbf{G}_z & - \mbf{G}_q  \\ \mbf{A}_p \mbf{G}_z & \mbf{0} \end{smallmatrix}\right], \left[\begin{smallmatrix} \mbf{b}_z \\ \mbf{c}_q - \mbf{H}_p \mbf{c}_z \\ \mbf{b}_p - \mbf{A}_p \mbf{c}_z \end{smallmatrix}\right] \right)\czon,
  \end{aligned}
	\end{equation}
	with $\mbf{G}_q \triangleq \half \text{diag} (\mbf{k}_p-\bm{\sigma})$, and $\mbf{c}_q \triangleq \half (\mbf{k}_p+\bm{\sigma})$ for any $\bm{\sigma} \in \realset^{n_h}$ satisfying $\bm{\sigma} \leq \mbf{H} \mbf{z},~ \forall \mbf{z} \in Z$.
\end{proposition}
\begin{proof} 
Let $P = P_\text{ineq} \cap P_\text{eq}$, where $P_\text{ineq} \triangleq (\mbf{H}_p, \mbf{k}_p, \noarg, \noarg)\poly$, and $P_\text{eq} \triangleq (\noarg, \noarg, \mbf{A}_p, \mbf{b}_p)\poly$. Then, it is true that $Z \cap P = Z \cap (P_\text{ineq} \cap P_\text{eq}) = (Z \cap P_\text{ineq}) \cap P_\text{eq}$. By the definition of $\bm{\sigma}$ and $P_\text{ineq}$, $\mbf{z} \in Z \cap P_\text{ineq} \implies \mbf{H}_p \mbf{z} \in [\bm{\sigma}, \mbf{k}_p]$. Let $Q \triangleq [\bm{\sigma}, \mbf{k}_p] = (\half \text{diag}(\mbf{k}_p - \bm{\sigma}), \half (\mbf{k}_p + \bm{\sigma}))\zon \triangleq (\mbf{G}_q,\mbf{c}_q)\zon \subset \realset^{n_h}$. Then, $Z \cap P_\text{ineq} = \{\mbf{z} \in Z: \mbf{H}_p \mbf{z} \leq \mbf{k}_p\} = \{\mbf{z} \in Z: \mbf{H}_p \mbf{z} \in Q \} = Z \cap_{\mbf{H}_p} Q$. Therefore, $\mbf{z} \in (Z \cap P_\text{ineq}) \cap P_\text{eq} \iff \mbf{z} \in (Z \cap_{\mbf{H}_p} Q) \cap P_\text{eq} \iff \exists (\bm{\xi}_z, \bm{\xi}_q) \in B_\infty(\mbf{A}_z,\mbf{b}_z) \times B_\infty^{n_h}$ such that
\begin{align}
\mbf{z} & = \mbf{c}_z + \mbf{G}_z \bm{\xi}_z, \label{eq:czhrepintersection_aux1} \\
\mbf{H}_p \mbf{z} & = \mbf{c}_q + \mbf{G}_q \bm{\xi}_q, \label{eq:czhrepintersection_aux2}\\
\mbf{A}_p \mbf{z} & = \mbf{b}_p. \label{eq:czhrepintersection_aux3}
\end{align}
Substituting \eqref{eq:czhrepintersection_aux1} into \eqref{eq:czhrepintersection_aux2} and \eqref{eq:czhrepintersection_aux3}, and gathering the resulting equations with \eqref{eq:czhrepintersection_aux1}, leads to the CG-rep \eqref{eq:czhrepintersection}.
\end{proof}

\subsection{Reachability Analysis of Nonlinear Systems}

Consider the nonlinear system \eqref{eq:systemf} and let $\mbf{f}$ be a factorable function with factors $\mbf{z}$ and $\mbf{E}_f \in \realsetmat{n_x}{n_z}$ satisfying $\mbf{f}(\mbf{x}_{k-1},\mbf{w}_{k-1}) = \mbf{E}_f\mbf{z}$. Assuming that $\mbf{x}_0 \in \bar{X}_0$ and $\mbf{w}_{k} \in W, ~\forall k \geq 0$, with $\bar{X}_0 \subset \realset^{n_x}$ and $W \subset \realset^{n_w}$ being constrained zonotopes, reachability analysis of \eqref{eq:systemf} is performed by recursively computing CZ enclosures $\bar{X}_k$ satisfying \eqref{eq:reachrecursive} using the developed propagation method \eqref{eq:czenclosurefinal}. Specifically, let $P_{k-1}\subset \realset^{n_z}$ be the halfspace polytope computed as in \eqref{eq:liftedpolyhedron} for the function $\mbf{f}$ on the interval $\square(\bar{X}_{k-1} \times W)$. 
Then, $\mbf{x}_k = \mbf{f}(\mbf{x}_{k-1},\mbf{w}_{k-1}) \in \bar{X}_k$, where
\begin{equation} \label{eq:czreachablerecursive}
\bar{X}_k \triangleq \mbf{E}_f ((\bar{X}_{k-1} \times W \times \tilde{Z}_{k-1}) \cap P_{k-1}).
\end{equation}

\begin{remark} \rm \label{rem:setcomplexity}
    Let $\bar{X}_{k-1}$ have $n_{g_{k-1}}$ generators, and $n_{c_{k-1}}$ constraints. Moreover, let each $P_{k-1}$ given by \eqref{eq:liftedpolyhedron} have $n_{h_p}$ halfspaces and $n_{c_p}$ equality constraints. Then, the CZ enclosure $\bar{X}_k$ obtained by \eqref{eq:czreachablerecursive} has $n_{g_k} = n_{g_{k-1}} + n_{g_w} + n_z - (n_x + n_w) + n_{h_p}$ generators and $n_{c_k} = n_{c_{k-1}} + n_{c_w} + n_{h_p} + n_{c_p}$ constraints. Note that $n_z$, $n_{h_p}$ and $n_{c_p}$ are constant values, which depend only on the (non-unique) factorization of $\mbf{f}$. Therefore, similar to the CZ-based mean value extension in \cite{Rego2021}, the complexity increase of the proposed CZ enclosure is linear. This is in contrast to the CZ-based first-order Taylor extension in the same reference, which has quadratic growth. It is also in contrast to classical polytope-based estimation methods, which suffer from exponential growth in the number of halfspaces required at each sampling time.
\end{remark}

\begin{remark} \rm \label{rem:dependencies}
    Although the dependencies between states are neglected when using $\square \bar{X}_{k-1}$ to compute $P_{k-1}$, these dependencies are recovered by the intersection in \eqref{eq:czreachablerecursive}.
\end{remark}

Algorithm \ref{alg:czreachability} summarizes the new method for reachability analysis of \eqref{eq:systemf} for a given time horizon $N \geq 0$. Due to the complexity increase observed in Remark \ref{rem:setcomplexity}, the polynomial-time complexity reduction methods for CZs proposed in \cite{Scott2016} are applied to $\bar{X}_k$ at the end of each time step $k$. This operation is denoted by $\text{red}(\cdot)$.

\begin{algorithm}[!htb] 
	\caption{Proposed Reachability analysis of \eqref{eq:systemf}.}
	\label{alg:czreachability}
	\small
	\begin{algorithmic}[1]
		\State Let $\mbf{x}_0 \in \bar{X}_0 \subset \realset^{n_x}$ and $\mbf{w}_k \in W \subset \realset^{n_w}$, where $\bar{X}_0$ and $W$ are constrained zonotopes. 
  	    \For {$k = 1,\ldots,N$}
        \State $X \gets \bar{X}_{k-1} \times W$;
        \State Compute the interval hull $\square X$;
        \State Compute $Z$ by natural interval extension of the factors $z_j$ using $\square X$;        
        \State Compute $Q_j$ for each factor $z_j$, $j>n_x+n_w$, with $Z$ as domain;
        \State $P \gets \bigcap_{j=n_x+n_w+1}^{n_z} Q_j$;
        \State Obtain $\tilde{Z}$ from $Z$;        
        \State $\bar{X}_k \gets \mbf{E}_f ((\bar{X}_{k-1} \times W \times \tilde{Z}) \cap P)$;
        \State $\bar{X}_k \gets \text{red}(\bar{X}_k)$;
        \EndFor
	\end{algorithmic}
	\normalsize
\end{algorithm}

\section{Numerical Examples} \label{sec:examples}

\begin{figure*}[!tb]
	\centering{\!\!\!\!\subfloat[]{
		\def\svgwidth{0.665\columnwidth}
  {\scriptsize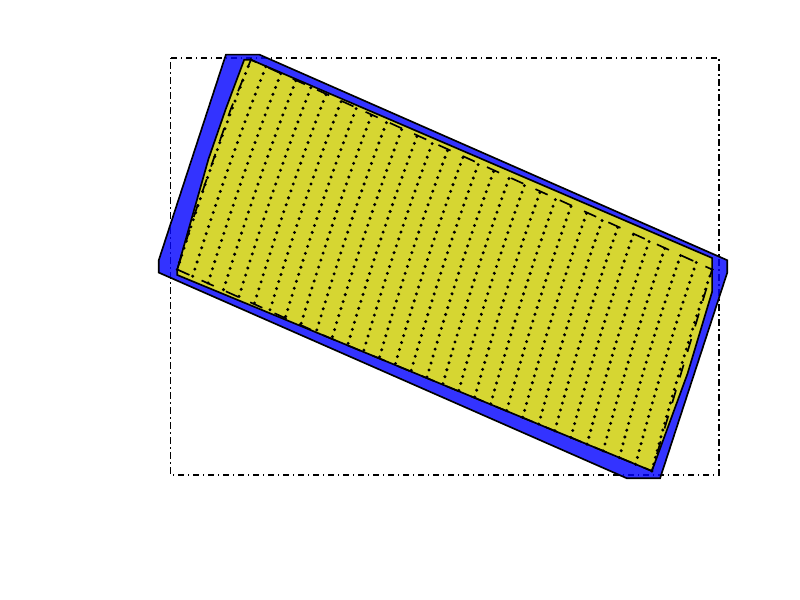}\label{subfig:nonlinearDCa}} 
  \subfloat[]{
		\def\svgwidth{0.665\columnwidth}
  {\scriptsize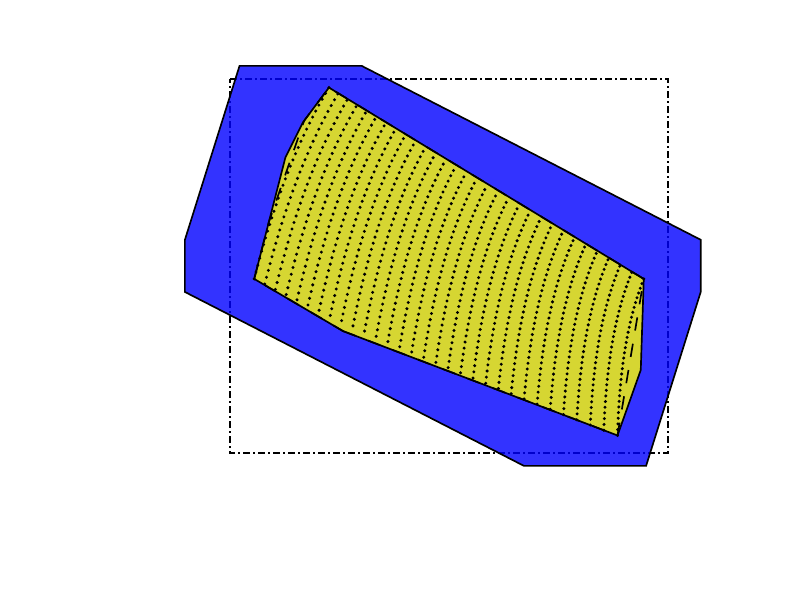}\label{subfig:nonlinearDCb}}
  \subfloat[]{
		\def\svgwidth{0.665\columnwidth}
  {\scriptsize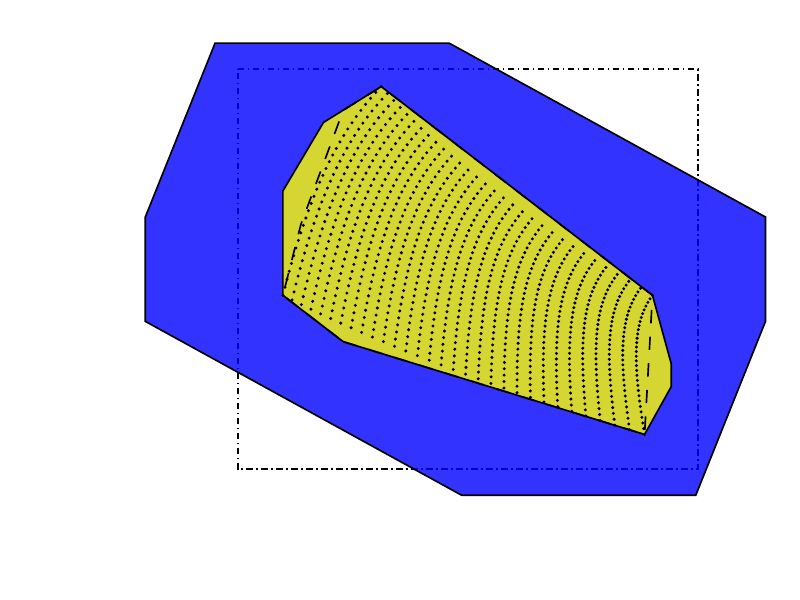}\label{subfig:nonlinearDCc}}} \\
  \centering{\!\!\!\!\subfloat[]{
		\def\svgwidth{0.665\columnwidth}
  {\scriptsize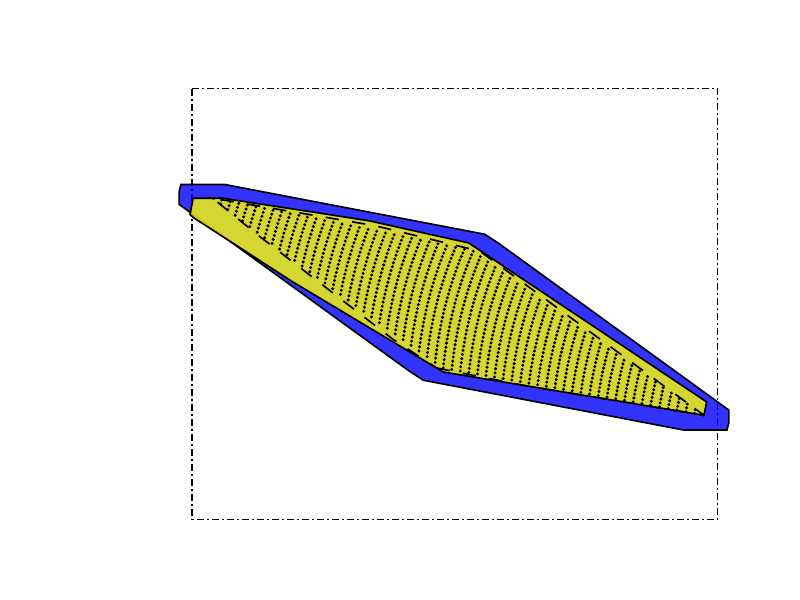}\label{subfig:nonlinearDCd}}
    \subfloat[]{
		\def\svgwidth{0.665\columnwidth}
  {\scriptsize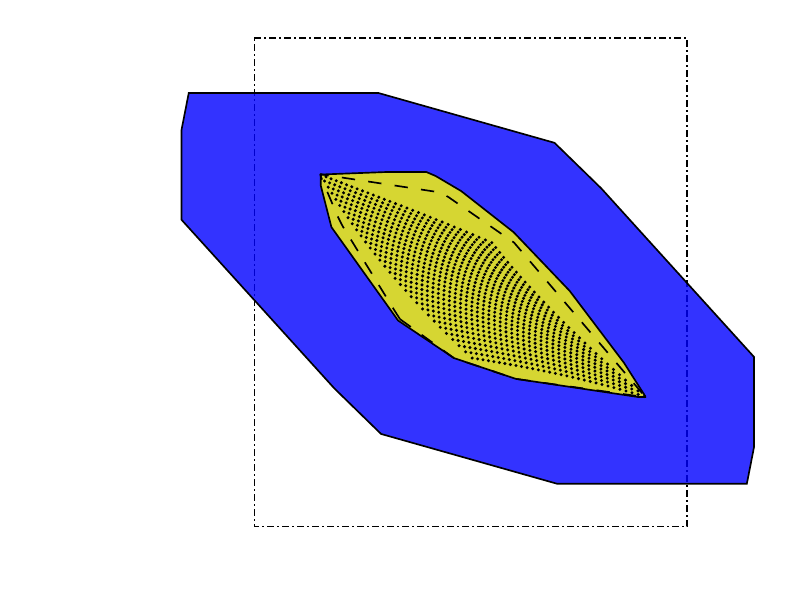}\label{subfig:nonlinearDCe}}
    \subfloat[]{
		\def\svgwidth{0.665\columnwidth}
  {\scriptsize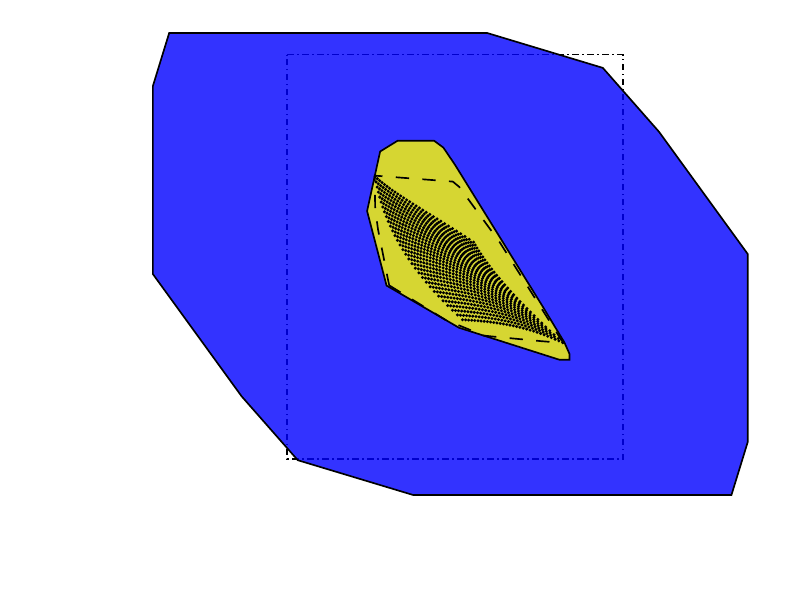}\label{subfig:nonlinearDCf}}}
  \caption{Enclosures for Example 1 obtained by IA (dash-dotted lines), CZMV (solid blue), Alg\ref{alg:czreachability} with complexity reduction (solid yellow), and Alg\ref{alg:czreachability} without complexity reduction (dashed lines), along with uniform samples from $\bar{X}_0$ propagated through \eqref{eq:nonlinearDC} (dots). Panels (a)--(c) show $\bar{X}_1$ for $\alpha = 0.1$, $0.5$, $1$, respectively, while panels (d)--(f) show $\bar{X}_2$ for $\alpha = 0.1$, $0.5$, $1$, respectively. This figure has been generated using the YALMIP function \emph{plot}.}\label{fig:nonlinearDC}
\end{figure*}

This section demonstrates the results obtained by the proposed CZ enclosure method (Algorithm \ref{alg:czreachability}, denoted as Alg\ref{alg:czreachability}), and compares with results obtained by applying interval arithmetic to $\mbf{f}$ (IA) \cite{Moore2009} and using the CZ-based mean value extension (CZMV) described in \cite{Rego2021}. Numerical simulations were performed using MATLAB 9.1 with LPs solved by Gurobi 10.0.1. For all methods, the number of constraints and generators in the CZs is limited to 20 and 8, respectively, using the reduction methods in \cite{Scott2016}. 

\subsection{Example 1}
\label{sec:example 1}
We first consider the computation of enclosures $\bar{X}_k \supseteq \{\mbf{f}(\mbf{x}): \mbf{x} \in \bar{X}_{k-1}\}$ with
\begin{equation} \label{eq:nonlinearDC}
\mbf{f}(\mbf{x}) \triangleq
\begin{bmatrix}
x_2(-0.7 + 0.1x_2 + 0.1x_1) + 0.1e^{x_1} \\
x_1(   1 - 0.1x_1 + 0.2x_2) + x_2
\end{bmatrix}.
\end{equation}
The initial domain is $\bar{X}_0 \triangleq ( \alpha \eye{2}, \zeros{2}{1})\zon$ with $\alpha \in (0,1]$. Fig.~\ref{fig:nonlinearDC} shows the enclosures obtained by IA, CZMV, and Alg\ref{alg:czreachability}, for $k=\{1,2\}$ and $\alpha \in \{0.1,0.5,1\}$, together with uniform samples from $\bar{X}_0$ propagated through \eqref{eq:nonlinearDC}. When $\bar{X}_0$ is small, Alg\ref{alg:czreachability} and CZMV provide similar enclosures, which are both more accurate than IA. However, with larger $\bar{X}_0$, the conservatism of CZMV can lead to worse enclosures than even IA, while Alg\ref{alg:czreachability} computes significantly better enclosures that approximate the propagated samples well.

\subsection{Example 2}
\label{sec:example 2}
We now consider the reachability analysis of an isothermal gas-phase reactor with dynamic equations discretized using the forward Euler method \cite{Alamo2005a,Rego2018}:
\begin{equation} \label{eq:isothermal}
\begin{aligned}
x_{1,k} & = x_{1,k-1} + T_s \left( -2k_1 x_{1,k-1}^2  + 2k_2 x_{2,k-1}\right), \\
x_{2,k} & = x_{2,k-1} + T_s \left( k_1 x_{1,k-1}^2  - k_2 x_{2,k-1}\right),
\end{aligned}
\end{equation}
where $k_1 = 0.16/60$ s$^{-1}$ atm$^{-1}$, $k_2 = 0.0064/60$ s$^{-1}$, and $T_s = 6$ s. The initial set is
\begin{equation*}
	\bar{X}_0 \triangleq \left( \begin{bmatrix} 2.5 & -0.2 & 0.1 \\ 0.5 & 0.5 & 0.1 \end{bmatrix}, \begin{bmatrix} 2.5 \\ 1 \end{bmatrix}, \begin{bmatrix} 1 & -0.1 & 1 \end{bmatrix}, 1 \right)\czon.
\end{equation*}
In this example, we also compare the results obtained by the CZ-based first-order Taylor extension described in \cite{Rego2021}, denoted as CZFO. For the analysis, we define the 1-norm radius (1-radius, in short) as $\text{rad}_1(\bar{X}_k) \triangleq \sum_{i=1}^{n_x} \rad{\zeta}_i$, where $\bm{\zeta} \triangleq \square \bar{X}_k \in \intvalset^{n_x}$.

\begin{figure}[!tb]
	\centering{
		\def\svgwidth{\columnwidth}
  {\scriptsize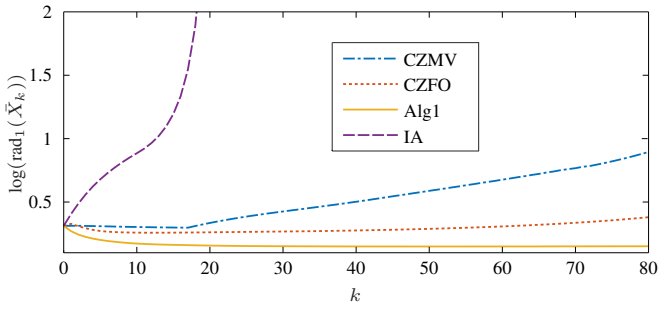}
		\caption{The 1-norm radii of the enclosures $\bar{X}_k$ obtained by IA (dashed purple), CZMV (dash-dotted blue), CZFO (dotted red), and Alg\ref{alg:czreachability} (solid yellow) for Example 2.}\label{fig:Alamo2005}}
\end{figure}

Fig.~\ref{fig:Alamo2005} shows the results obtained by IA, CZMV, CZFO, and Alg\ref{alg:czreachability}. The interval enclosures generated by IA diverge after a few times steps, while the enclosures generated by Alg\ref{alg:czreachability} are less conservative than those generated by CZMV and CZFO. Although CZFO produces reasonable enclosures, it exhibits a quadratic increase in the number of generators and constraints in each time step \cite{Rego2021}, whereas both CZMV and Alg\ref{alg:czreachability} exhibit linear increases. This leads to increased cost in the reduction step. In particular, once the maximum allowable complexity is reached, the reduction step for CZFO must eliminate 212 generators and 36 constraints from $\hat{X}_k$ in each step, compared to only 20 generators and 18 constraints for Alg\ref{alg:czreachability}, and only 2 generators and zero constraints for CZMV (in this case, the obtained enclosure had the same number of constraints as the previous CZ). On a laptop with 32 GB RAM and an Intel Core i7-12700H processor, the average computational times of CZMV, CZFO, and Alg\ref{alg:czreachability}, at each $k$, were 6.4 ms, 23.5 ms, and 25 ms, respectively.

\section{Conclusions} \label{sec:conclusions}

This paper proposed a novel algorithm for reachability analysis of nonlinear discrete-time systems using constrained zonotope enclosures. By combining important properties of CZs and polyhedral relaxations of factorable representations of nonlinar functions, the new method was able to generate outer-approximations that are less conservative than the ones obtained by interval methods and other CZ methods from the literature, while having only a linear complexity increase in the computed enclosures.

\bibliography{\bibfolder/masterthesis_bib,\bibfolder/appendices_bib,\bibfolder/UAVControl_bib,\bibfolder/BackgroundHist_bib,\bibfolder/Surveys_bib,\bibfolder/PassiveFTC_bib,\bibfolder/ActiveFTC_bib,\bibfolder/UAVFTC,\bibfolder/SetTheoretic_bib,\bibfolder/SetTheoreticFTCFDI_bib,\bibfolder/Davide_bib,\bibfolder/paperAutomatica_bib,\bibfolder/paperCDC_bib,\bibfolder/paperECC_bib,\bibfolder/paperIFAC_bib,\bibfolder/paperNonlinearMeas_bib,\bibfolder/Robotic_bib,\bibfolder/stelios_bibliography,\bibfolder/phdthesis_bib,\bibfolder/paperParameter_bib,\bibfolder/Diego_bib,\bibfolder/paperMixed_bib,\bibfolder/paperMcCormick_bib}

\end{document}